\documentclass{ijmpb}
\begin{document}
\title{MODIFIED JOSEPHSON RELATION}
\author{Jan Kol\'a\v cek, Pavel Lipavsk\'y}
\address{Institute of Physics, Academy of Sciences, 
Cukrovarnick\'a 10, 16253 Prague 6, Czech Republic}
\maketitle

\begin{abstract}

For type II superconductors, Josephson has shown that vortices 
moving with velocity ${\bf v}_{\rm L}$ create an {\em effective} 
electric field  ${\bf  E}'=-{\bf v}_{\rm L}\times{\bf B}_{\rm V}$. 
By definition
the effective electric field is gradient of the electrochemical
potential, what is the quantity corresponding to voltage observed 
with the use of Ohmic contacts. It relates to the true 
electric field ${\bf E}$ via the local chemical potential 
$\mu$ as ${\bf E}'={\bf E}-\nabla\mu/e$. 
We argue that at low temperatures the true electric field 
in the bulk can be approximated by a modified 
Josephson relation ${\bf E}=({\bf v}_{\rm S}-{\bf v}_{\rm L} )\times
{\bf B}_{\rm V}$, where ${\bf v}_{\rm S}$ is the condensate velocity. 
\end{abstract}


\section{Introduction}

For type II superconductors with moving vortices 
Josephson\cite{65Josephson} derived the formula 
\begin{equation}
{\bf  E}'\equiv -\nabla(\varphi+\mu/e)-\partial{\bf A}/\partial t 
          = - {\bf v}_{\rm L} \times {\bf B}_{\rm V}.
\label{J}
\end{equation}
The effective electric field ${\bf  E}'$ is a gauge-invariant 
generalization of the gradient of the electrochemical potential
$\varphi+\mu/e$, with $\mu$ being the local chemical 
potential,\footnote{Here we are using the nowadays commonly
used convention in which the so called electrochemical potential is 
identical with the Gibbs chemical potential $\mu_{\rm Gibbs}=\mu+e\varphi$. 
With a good approximation, the electrostatic potential $\varphi$ 
represents the long-range interaction while the chemical potential $\mu$ 
is a local functional of the density, temperature, BCS gap, and 
so on. Josephson in his original paper\cite{65Josephson} 
uses the term chemical potential for the Gibbs chemical potential.} 
${\bf v}_{\rm L}$ is the velocity of the vortex lattice and 
${\bf B}_{\rm V}=n_{\rm V}{\bf\phi}_0$ 
is the averaged magnetic field generated by vortices distributed 
with the density~$n_{\rm V}$. 

The Josephson relation (\ref{J}) proved to be useful in situations 
when one evaluates the difference between electrochemical potentials 
as measured by Ohmic contacts. On the other hand it is not explaining 
the voltage measured by a 
contactless capacitive pickup\cite{68BokKlein,71MorrisBrown}.
It also cannot be used for interpretation of the FIR data,\cite{HFvordyn} 
because electric field of the laser light has to be matched with the 
true electric field ${\bf E}=-\nabla\varphi-\partial{\bf A}/\partial t$,
not with its effective counterpart ${\bf E}'$.

Vector potential is the linear and electrostatic potential 
is the quadratic function of the local condensate velocity\cite{50London}, 
so in principle one can evaluate the electric field directly from its 
definition. Such approach would require a microscopic picture of the moving 
Abrikosov vortex lattice what represents a demanding numerical task. 
We argue that at low temperatures, when the contribution of normal
electrons can be neglected, the true electric field can be
approximated as 
\begin{equation}
{\bf E}=\left({\bf v}_{\rm S}-{\bf v}_{\rm L}\right)\times
{\bf B}_{\rm V},
\label{mJ}
\end{equation}
where ${\bf v}_{\rm S}$ is the mean velocity of condensate obtained
by averaging over the elementary cell of the Abrikosov vortex lattice. 

\section{Electric field in type I superconductor}

It was theoretically predicted\cite{50London} and experimentally 
confirmed\cite{68BokKlein,71MorrisBrown} that superconducting 
current generates transversal electric field. In the planar 
geometry, this field can be derived from the Newton equation
$m\dot{\bf v}_{\rm S}=e{\bf E}+e{\bf v}_{\rm S}\times{\bf B}$. Since the
acceleration is zero, one finds ${\bf E}=-{\bf v}_{\rm S}\times{\bf B}$.
From the London condition $\nabla\times{\bf v}_{\rm S}=-e{\bf B}/m$
solved by $m{\bf v}_{\rm S}=-e{\bf A}$ one can derive a more general
formula $e\varphi=-{1\over 2}m{\bf v}_{\rm S}^2$, which applies also for 
curved streamlines with nonzero centripetal acceleration and for time
dependent vector potentials. Electric field is acting also on 
the crystal lattice and by this way the Lorentz force is mediated to it. 

At finite temperatures the presence of quasiparticles must be also 
taken into account. When flowing, normal electrons dissipate energy
and therefore, in spite of the presence of an electric field,
the average velocity of the normal state fluid ${\bf v}_{\rm N}$ 
is zero at equilibrium. This is explained by the interaction 
between the superconducting and the normal state fluid, known as 
quasiparticle screening\cite{VS64}.
Due to it the electrostatic Bernoulli potential is reduced by 
the share of the condensate so that it reads 
$e\varphi = -n_{\rm S}/n \left({1\over 2} m^*{\bf v}_{\rm S}^2 \right)$. 
Arguments explaining electric field in superconductors
can be found in the BCS theory\cite{AW68} and 
the extended Ginzburg-Landau theory\cite{02ElstPot}, too. 
The fact that electric field generated by surface currents 
mediate Lorentz force to the crystal lattice follows from 
the Budd-Vannimenus (BV) theorem\cite{BVtheorem}.
For simplicity in the following we restrict ourselves to low temperatures, 
where interaction with quasiparticles can be neglected. 
  
\section{Electric field in type II superconductor}

Superconducting current acts on the vortices by the Magnus force, which 
(supposing that vortices are aligned parallel to the $z$-axis) 
reads\cite{93Ao}
\begin{equation}
  {\bf F}_M^{\rm V}={n_{\rm S} h\over 2}
  \left( {\bf v}_{\rm S} - {\bf v}_{\rm L}\right)\times {\bf z}.  
\label{Magnus}
\end{equation}
It follows from the third Newton law that with the Magnus force 
acting on vortices also a reaction force felt by the superconducting particles
${\bf F}_M^{\rm S}=-{n_{\rm V} \over n_{\rm S}} {\bf F}_M^{\rm V}
     = -e\left( {\bf v}_{\rm S} - {\bf v}_{\rm L} \right)
       \times {\bf B}_{\rm V} $
must be introduced.

In steady state in which vortices and superconducting fluid move 
without acceleration, the total forces acting on them have to 
vanish. The superconducting particles feel the effect of 
electric field, Lorentz force (interaction with external
magnetic field $\bf B_{\rm ext}$ penetrating into the superconductor)  
and the reaction of Magnus force. 
From the Newton equation of motion 
$m\dot{\bf v}_{\rm S}=e{\bf E}
   +e{\bf v}_{\rm S}\times{\bf B}_{\rm ext}
   -e\left({\bf v}_{\rm S} -{\bf v}_{\rm L}\right)\times{\bf B}_{\rm V} $
one gets 
\begin{equation}
{\bf E} = 
   -{\bf v}_{\rm S}\times{\bf B}_{\rm ext}
   +\left({\bf v}_{\rm S} -{\bf v}_{\rm L}\right)\times{\bf B}_{\rm V}.
\label{mJe}
\end{equation} 
In the bulk, where external magnetic field is screened off,
the electric field can be approximated by  
modified Josephson relation (\ref{mJ}).
	
From this it follows, that the true electric field is nonzero even
if vortices are pinned. It is understandable.  
The sum of Magnus forces acting on the individual vortices is equal 
to the Lorentz force, which the current feels in magnetic field 
${\bf B}_{\rm V}$ created by the vortex lattice. 
Nevertheless, the Magnus force represents the interaction of 
the superconducting fluid with vortices, and not an interaction
with the external magnetic field ${\bf B}_{\rm ext}$. 
Pinned vortices can remain in the superconductor even if 
external magnetic field is switched off and in this case the 
total force acting on the crystal lattice is zero. 
Vortices are not moving, as they are kept by the pinning force.
The reaction of the Magnus force felt by the superfluid
is balanced by electric field (\ref{mJ}) and it also balances
pinning force felt by the crystal lattice. 
If external magnetic field is present, according to the BV theorem, 
electric field induced by the surface currents mediates
Lorentz force on the crystal lattice.

\section{Conclusion}
Effective electric field in type II superconductor is given 
by Josephson relation~(\ref{J}). 
At low temperatures in steady state the true electric field 
in the bulk can be approximated by the modified Josephson relation (\ref{mJ}).
In the presence of transport current, the Lorentz force acting on 
the superconducting wire in external magnetic field
is mediated by electric field generated by the surface currents. 
 
\section*{Acknowledgments}
This work was supported by M\v{S}MT program Kontakt ME 601 
and GA\v{C}R 202000643, GAAV A1010312 grants. 
The European ESF program VORTEX is also acknowledged.

\end{document}